\documentclass[showpacs,eqsecnum,aps]{revtex4}
\usepackage{amsmath}
\usepackage{amssymb}
\usepackage{amsfonts}
\def\beq{\begin{equation}}
\def\eeq{\end{equation}}
\def\beqa{\begin{eqnarray}}
\def\eeqa{\end{eqnarray}}

\def\sumint{\sum\mspace{-25mu}\int}

\def\sumint{\sum\mspace{-25mu}\int}

\usepackage[dvips]{graphicx,color}

\begin{document}

\title{Comment on the equivalence of Bakamjian-Thomas mass operators 
in different forms of dynamics}

\author{W. N. Polyzou}
\affiliation{
Department of Physics and Astronomy, The University of Iowa, Iowa City, IA
52242}

\vspace{10mm}
\date{\today}

\begin{abstract}

  We discuss the scattering equivalence of the generalized
  Bakamjian-Thomas construction of dynamical representations of the
  Poincar\'e group in all of Dirac's forms of dynamics.  The
  equivalence was established by Sokolov in the context of proving
  that the equivalence holds for models that satisfy cluster
  separability.  The generalized Bakamjian Thomas construction is used
  in most applications, even though it only satisfies cluster
  properties for systems of less than four particles.  Different forms
  of dynamics are related by unitary transformations that remove
  interactions from some infinitesimal generators and introduce them
  to other generators.  These unitary transformation must be
  interaction dependent, because they can be applied to a
  non-interacting generator and produce an interacting generator.
  This suggests that these transformations can generate complex
  many-body forces when used in many-body problems.  It turns out that
  this is not the case.  In all cases of interest the result of
  applying the unitary scattering equivalence results in
  representations that have simple relations, even though the unitary
  transformations are dynamical.  This applies to many-body models as
  well as models with particle production.  In all cases no new
  many-body operators are generated by the unitary scattering
  equivalences relating the different forms of dynamics.  
  This makes it clear that the various calculations used
  in applications that emphasize one form of the dynamics over another
  are equivalent.  Furthermore, explicit representations of the 
  equivalent dynamical models in any form of dynamics are easily
  constructed.  Where differences do appear is when electromagnetic
  probes are treated in the one-photon exchange approximation.  This
  approximation is different in each of Dirac's forms of dynamics.
 
\end{abstract}

\vspace{10mm}

\pacs{11.80.-m, 24.10.jv}

\maketitle

%\newpage

%****************************************************************************

%\narrowtext

%****************************************************************************

\section{Introduction}

One of the most straightforward constructions of exactly Poincar\'e
invariant quantum mechanical models of systems of a finite number of
degrees of freedom is based on a method introduced by Bakamjian and
Thomas \cite{Bakamjian:1953kh}.  The construction can be summarized as
follows.  Particles are represented by irreducible representations of
the Poincar\'e group.  The model Hilbert space, which is determined by
the particle content of the system, is the direct sum of tensor
products of irreducible representation spaces for the Poincar\'e
group.  The kinematic (non-interacting) unitary representation of the
Poincar\'e group, $U_0(\Lambda ,a)$, on this space is the direct sum
of tensor products of unitary irreducible representations of the
Poincar\'e group.  The kinematic representation of the Poincar\'e
group is decomposed into a direct integral of irreducible
representations of the Poincar\'e group using Poincar\'e group
Clebsch-Gordan coefficients\cite{Joos:1962}\cite{Coester:1965zz}
\cite{Keister:1991sb}.  Wave functions in this direct integral
representation are square integrable functions of the eigenvalues of
(1) the Casimir operators, $(m,j)$, of the Poincar\'e group (2)
commuting observables, $\mathbf{v}$, that label different vectors in
an irreducible subspace and (3) invariant degeneracy operators,
$\mathbf{d}$, that distinguish multiple copies of the same irreducible
representation.  Wave functions in this representation are square
integrable functions, $\psi (m,j,\mathbf{v},\mathbf{d})$, of the eigenvalues
of these operators.

The goal of the Bakamjian-Thomas construction is to add interactions
to the Poincar\'e generators in a manner that preserves the Poincar\'e
Lie algebra.  This is non-trivial because the Hamiltonian appears on
the right side of the commutator of the translation and boost generators,
\beq
[P^j ,K^k] = i \delta_{jk}H ,  
\label{a.1}
\eeq
which cannot be satisfied for an interacting $H$ unless some
combination of $\mathbf{P}$ and $\mathbf{K}$ also include
interactions.  The full set of commutation relations imposes
additional non-linear constraints on the interactions.

Bakamjian and Thomas solve this problem by adding interactions to the
mass Casimir operator, $m$.  The allowed interactions are
represented by kernels that have the form,
\beq
\langle (m,j), \mathbf{v}, \mathbf{d} \vert V \vert (m',j'), \mathbf{v}',
\mathbf{d}' \rangle =
\delta (\mathbf{v}:\mathbf{v}') \delta_{jj'} \langle m, \mathbf{d} \Vert V^j \Vert m',
\mathbf{d}'\rangle 
\label{a.2}
\eeq
in the kinematic irreducible representation, where $\delta
(\mathbf{v}:\mathbf{v})$ denotes a product of Dirac delta functions in
the continuous variables and Kronecker delta functions in the discrete
variables.  If $m_d=m_d^{\dagger}:=m+V >0$ then $m_d$ becomes the mass Casimir
operator for a dynamical representation of the Poincar\'e group.  The
structure of the interaction and the requirement $m_d>0$ implies that
simultaneous eigenstates of $m_d$, $j^2$ and $\mathbf{v}$, denoted by
$\vert (\lambda,j), \mathbf{v} \rangle$, are complete, and transform
irreducibly with respect to a dynamical representation of the
Poincar\'e group.  Simultaneous eigenfunctions of $\{m_d,j,\mathbf{v} \}$ in the
kinematic irreducible basis have the form
\beq 
\langle
(m,j), \mathbf{v}, \mathbf{d} \vert (\lambda',j'), \mathbf{v}' \rangle
= \delta (\mathbf{v}:\mathbf{v}') \delta_{jj'} \psi_{\lambda' ,j'}
(m,\mathbf{d}) 
\label{a.3}
\eeq 
where the internal wave-function, $\psi_{\lambda',j'} 
(m,\mathbf{d})$, is the solution of the eigenvalue equation:
\beq 
(\lambda -m) \psi_{\lambda' ,j'} (m,\mathbf{d}) = \sumint{}'
\langle m, \mathbf{d} \Vert V^j \Vert m', \mathbf{d}'\rangle dm'
d\mathbf{d}' \psi_{\lambda' ,j'} (m',\mathbf{d}').  
\label{a.4}
\eeq 
Note that the variables $\mathbf{v}$, which define the choice of basis 
on each irreducible subspace,  do not appear in the equation for
the internal wave function, $\psi_{\lambda' ,j'} (m',\mathbf{d}')$.  
In addition, the variables $\mathbf{v}$ play no role in formulating 
the asymptotic conditions for scattering solutions of equation
(\ref{a.4}).

This means that the internal wave function $\psi_{\lambda' ,j'}
(m,\mathbf{d})$ is independent of the choice of basis for the
kinematic irreducible representation.  The dynamical unitary 
representation of the Poincar\'e group on this 
complete set of eigenstates is
\beq
\langle  (m,j), \mathbf{v}, \mathbf{d} \vert 
U(\Lambda ,a) 
\vert
(\lambda,j), \mathbf{v}' \rangle =
\sumint{}'' 
\langle  (m,j), \mathbf{v}, \mathbf{d} \vert (\lambda,j), \mathbf{v}'' \rangle
d\mathbf{v}''
{\cal D}^{\lambda ,j}_{\mathbf{v}'',\mathbf{v}'}[\Lambda ,a]
\label{a.5}
\eeq
where 
\beq
{\cal D}^{\lambda ,j}_{\mathbf{v}'',\mathbf{v}'}[\Lambda ,a] :=
\langle  (\lambda,j), \mathbf{v}''  \vert U(\Lambda ,a) 
\vert (\lambda,j), \mathbf{v}' \rangle
\label{a.6}
\eeq
is the Poincar\'e group Wigner function, which is the known 
mass $\lambda$ spin $j$ irreducible representation of the Poincar\'e group 
in the basis $\{\vert (\lambda,j), \mathbf{v}' \rangle\}$, which 
we call the ``$\mathbf{v}$-basis''.   The Wigner function is 
dynamical because it depends on the mass eigenvalue $\lambda$, 
which requires solving eq. (\ref{a.4}).
%but the mass spectrum only depends on the reduced interaction 
%$\langle m, \mathbf{d} \Vert V^j \Vert m',
%\mathbf{d}'\rangle$. 

This is a short summary of the Bakamjian-Thomas construction.  This
construction gives an explicit representation of finite Poincar\'e
transformations.  Dynamical generators can be constructed by
differentiating with respect to the group parameters.  Bakamjian and
Thomas actually construct the generators, but they are difficult to
exponentiate, while the finite transformations discussed above can be
used directly in applications.
  
The Bakamjian-Thomas construction
is not limited to two-particle or fixed number of particle systems.  
In more complex systems the
interaction is a sum of interactions that may be more naturally
expressed in bases with the same $\mathbf{v}$ but different degeneracy
parameters.  For example, in the three-body problem it is natural to
construct three-body kinematic irreducible representation using
successive pairwise coupling.  Different orders of pairwise coupling
lead to irreducible representations with the same overall
$\mathbf{v}$ but different choices of degeneracy parameters,
$\mathbf{d}$.  For example, interactions involving the $i-j$ pair of particles are
most naturally described in a representation where the $i-j$ pair are
coupled first.

Because the degeneracy parameters are kinematically invariant, the
coefficients of the transformation that relates bases with degeneracy
parameters $\mathbf{d}_b$ to bases with degeneracy parameters
$\mathbf{d}_a$ necessarily have the form
\beq
\langle  (m,j), \mathbf{v}, \mathbf{d}_a \vert  
(m',j'), \mathbf{v}', \mathbf{d}_b' \rangle = \delta (\mathbf{v}:\mathbf{v}')
\delta_{jj'} \delta (m:m')  
R^{jm}(\mathbf{d}_a,\mathbf{d}_b') .
\label{a.7}
\eeq
The coefficients $R^{jm}(\mathbf{d}_a,\mathbf{d}_b')$ of the unitary 
operator that transforms invariant degeneracy parameters are Racah 
coefficients for the Poincar\'e group.  The important observation is
that these coefficients commute with and are independent of the 
variables, $\mathbf{v}$.

In the general case the interaction kernel, (\ref{a.2}),  has the form 
\[
\langle (m,j), \mathbf{v}, \mathbf{d} \vert V \vert (m',j'), \mathbf{v}',
\mathbf{d}' \rangle =
\]
\beq
= \delta (\mathbf{v}:\mathbf{v}') \delta_{jj'} \sumint 
R^{jm}(\mathbf{d},\mathbf{d}_b) d\mathbf{d}_b
\langle m, \mathbf{d}_b \Vert V_b^j \Vert m', \mathbf{d}_b'\rangle
d\mathbf{d}_b'
R^{jm}(\mathbf{d}_b',\mathbf{d}') .
\label{a.8}
\eeq
The relevant observation is that in general the interaction still has
the form (\ref{a.2}) with 
\beq
\langle m, \mathbf{d} \Vert V^j \Vert m',
\mathbf{d}'\rangle =
\sumint 
R^{jm}(\mathbf{d},\mathbf{d}_b) d\mathbf{d}_b
\langle m, \mathbf{d}_b \Vert V_b^j \Vert m', \mathbf{d}_b'\rangle
d\mathbf{d}_b'
R^{jm}(\mathbf{d}_b',\mathbf{d}') .
\label{a.9}
\eeq

To make the connection with Dirac's forms of dynamics note that for
some choice of bases, $\vert (m,j)\mathbf{v}, \mathbf{d} \rangle$, the
Poincar\'e group Wigner function ${\cal D}^{\lambda
  ,j}_{\mathbf{v}'',\mathbf{v}'}[\Lambda ,a]$ is independent of the
mass $\lambda$ when $(\Lambda ,a)$ is restricted to a subgroup of the
Poincar\'e group.  The kinematic subgroup only depends on the choice
of basis, $\mathbf{v}$.  This is because the Poincar\'e group Wigner
function does not depend on the degeneracy parameters, $\mathbf{d}$.
This subgroup is called the kinematic subgroup associated with the
basis $\mathbf{v}$.  Dirac identified the three largest kinematic
subgroups, which are the three-dimensional Euclidean group
(instant-form dynamics), the Lorentz group (point-form dynamics), and
the subgroup that leaves a plane tangent to the light-cone invariant
(front-form dynamics).  In our presentation, each kinematic subgroup
is uniquely associated with a preferred basis for irreducible
subspaces.  This characterization exists even in the absence of
interactions.
  
The natural bases for the irreducible subspaces associated with 
Dirac's \cite{Dirac:1949cp} forms of dynamics are simultaneous eigenstates of
\begin{center}
Table 1.
\end{center}
\begin{center}
\begin{tabular}{lll}
%\begin{table}{lll}
\hline
form &  & vector variables \\
\hline
instant form: &$\mathbf{v} \to$ & ($\mathbf{p},\mathbf{j}_c\cdot \hat{\mathbf{z}})$\\
point form: &$\mathbf{v} \to$ & $ (\mathbf{u}:=\mathbf{p}/m,\mathbf{j}_c\cdot 
\hat{\mathbf{z}}) $\\
front form: &$\mathbf{v} \to$ & $ (p^+:=p^0+p^3,p^1,p^2,\mathbf{j}_f\cdot 
\hat{\mathbf{z}})  $ \\
%\end{table}
\hline
\end{tabular}
\end{center} 
where the $\mathbf{p}$ are momentum operators, $p^0$ is the 
Hamiltonian and $\mathbf{j}_x$ are different spin operators, 
which are related by momentum-dependent rotations\cite{Melosh:1974cu}.

The connection with Dirac's notion of kinematic subgroup is that when
$(\Lambda,a)$ is an element of the kinematic subgroup then
$U(\Lambda,a)$ can either act to the right on the parameters or 
to the left on the arguments of the wave function:
\[
\langle  (m,j) \mathbf{v}, \mathbf{d} \vert 
U[\Lambda ,a] 
\vert
(\lambda,j) \mathbf{v}' \rangle 
\]
\beq
=
\sumint{}''
{\cal D}^{m ,j}_{\mathbf{v},\mathbf{v}''}[\Lambda ,a] d\mathbf{v}''
\langle  (m,j) \mathbf{v}'', \mathbf{d}
\vert
(\lambda,j) \mathbf{v}' \rangle =
\sumint'' 
\langle  (m,j) \mathbf{v}, \mathbf{d} 
\vert
(\lambda,j) \mathbf{v}'' \rangle d\mathbf{v}''
{\cal D}^{\lambda ,j}_{\mathbf{v}'',\mathbf{v}'}[\Lambda ,a].
\label{a.10}
\eeq
In this case ${\cal D}^{\lambda ,j}_{\mathbf{v}'',\mathbf{v}'}[\Lambda ,a]=
{\cal D}^{m ,j}_{\mathbf{v}'',\mathbf{v}'}[\Lambda ,a]$ because the Wigner 
functions are independent of $m$ or $\lambda$ for $(\Lambda,a)$
kinematic.

Thus, while the computation of a general Poincar\'e transformation,
\beq
\langle  (m,j) \mathbf{v}, \mathbf{d} \vert U[\Lambda ,a] \vert \Psi \rangle 
= 
\sumint{}' \psi_{\lambda ,j} (m,\mathbf{d})
{\cal D}^{\lambda ,j}_{\mathbf{v},\mathbf{v}'}[\Lambda ,a]
\psi^*_{\lambda ,j} (m',\mathbf{d}') dm' d\mathbf{v}'d\mathbf{d}'
\langle (m',j) \mathbf{v}', \mathbf{d}' \vert \Psi \rangle
\label{a.11}
\eeq
requires solutions of the eigenvalue problem (\ref{a.4}), for
$(\Lambda ,a)$ in the kinematic subgroup, we get an equivalent, but simpler 
result that does not require solutions of (\ref{a.4}):
\beq
\langle  (m,j) \mathbf{v}, \mathbf{d} \vert U[\Lambda ,a] \vert \Psi \rangle
=
\sumint d\mathbf{v}' 
{\cal D}^{m ,j}_{\mathbf{v},\mathbf{v}'}[\Lambda ,a]
\langle (m',j) \mathbf{v}', \mathbf{d}' \vert \Psi \rangle .
\label{a.11}
\eeq

\section{The equivalence}

We call two theories scattering equivalent if (1) the unitary
representations of the Poincar\'e group are related by a unitary
transformation and (2) both theories have the same $S$-matrix
elements.  Note the that the first condition does not imply the
second.  Two-body models with different repulsive potentials are 
unitarily equivalent,
but they do not necessarily have identical phase shifts.

We consider a model Hilbert space defined by finite direct sums of
tensor products of irreducible representations.  We consider two
different single particle bases: $\vert (m,j) \mathbf{v}_a \rangle$
and $\vert (m,j) \mathbf{v}_b \rangle$.  These could be any pair of
bases from the table above or more generally the $\mathbf{v}$ could be
any set of observables that label vectors in an irreducible subspace.
In all cases these kinematic bases are related by a matrix for the
form
\beq
\langle (m,j) \mathbf{v}_a \vert (m',j') \mathbf{v}'_b \rangle =
\delta_{mm'}\delta_{jj'}A^{mj}(\mathbf{v}_a;\mathbf{v}'_b) 
\label{b.1}
\eeq

The kinematic irreducible bases are constructed out of direct sums of
tensor products of single-particle irreducible representations.  For
our purposes it is enough to consider successive pairwise coupling.
The coefficients of the unitary transformation relating tensor
products in the $a$ (resp $b$) basis to irreducible representation in
the $a$ (resp $b$) basis are Clebsch-Gordan coefficients of the
Poincar\'e group:

\[
\langle (m_1,j_1), \mathbf{v}_{a1}, (m_2,j_2)  \mathbf{v}_{a2}  
\vert (m,j), \mathbf{v}_a, \mathbf{d}  \rangle .
\]
These coefficients have the intertwining property
\[
\int \langle (m_1,j_1), \mathbf{v}_{a1}, (m_2,j_2)  \mathbf{v}_{a2}  
\vert (m',j'), \mathbf{v}''_a, \mathbf{d}''  \rangle 
d\mathbf{v}''
{\cal D}^{m' ,j'}_{\mathbf{v}'',\mathbf{v}'}[\Lambda ,a] =
\]
\beq
\int'' \prod {\cal D}^{m_1 ,j_1}_{\mathbf{v}_{a1},\mathbf{v}''_{a1}}
[\Lambda ,a]
{\cal D}^{m_2 ,j_2}_{\mathbf{v}_{a2},\mathbf{v}''_{a2}}
[\Lambda ,a]
d\mathbf{v}_{a1}''d\mathbf{v}_{a2}''
\langle (m_1,j_1), \mathbf{v}_{a1}'', (m_2,j_1) \mathbf{v}_{a2}''  
\vert (m,j) ,\mathbf{v}', \mathbf{d} \rangle .
\label{b.2}
\eeq
Depending on details of the construction of the Clebsch-Gordan 
coefficients there are different possible choices of degeneracy quantum 
numbers, $\mathbf{d}$.  

Keeping $\mathbf{d}$ fixed we can construct a Clebsch-Gordan coefficient in the 
$b$ basis using 
\[
\langle (m_1,j_1), \mathbf{v}_{b1}, (m_2,j_2)  \mathbf{v}_{b2}  
\vert (m,j), \mathbf{v}_b, \mathbf{d}  \rangle 
\]
\beq
= \sumint{}' d\mathbf{v}_{a1}'d\mathbf{v}_{a2}' d\mathbf{v}_{a}'
A^{m_1j_1}(\mathbf{v}_{b1};\mathbf{v}'_{a1}) 
A^{m_2j_2}(\mathbf{v}_{b2};\mathbf{v}'_{a2}) 
\langle (m_1,j_1), \mathbf{v}_{a1}', (m_2,j_2)  \mathbf{v}_{a2}'  
\vert (m,j), \mathbf{v}_a', \mathbf{d}  \rangle 
A^{mj}(\mathbf{v}_a';\mathbf{v}_b).
\label{b.3}
\eeq
What is relevant is that this is a Clebsch Gordan coefficient in the 
$b$ basis with the same degeneracy parameters as the original 
one in the $a$ basis.  

We can continue successively pairwise coupling until the entire Hilbert space
is represented by a direct integral of irreducible representations 
in the $a$ or $b$ basis with {\it identical degeneracy parameters} $\mathbf{d}$.
We write these bases as
\beq
\vert (m,j) \mathbf{v}_a, \mathbf{d} \rangle 
\qquad
\vert (m,j) \mathbf{v}_b, \mathbf{d} \rangle . 
\label{b.4}
\eeq   
What Sokolov established \cite{Sokolov:1977im}  was that 
the Bakamjian Thomas construction using the interactions
\beq
\langle (m,j), \mathbf{v}_a, \mathbf{d} \vert V_a \vert (m',j'), \mathbf{v}_b',
\mathbf{d}' \rangle =
\delta (\mathbf{v}_a:\mathbf{v}_a') \delta_{jj'} \langle m, \mathbf{d} \Vert V^j \Vert m',
\mathbf{d}'\rangle 
\label{b.5}
\eeq
and
\beq
\langle (m,j), \mathbf{v}_b, \mathbf{d} \vert V_b \vert (m',j'), \mathbf{v}_b',
\mathbf{d}' \rangle =
\delta (\mathbf{v}_b:\mathbf{v}_b') \delta_{jj'} \langle m, \mathbf{d} \Vert V^j \Vert m',
\mathbf{d}'\rangle 
\label{b.6}
\eeq
are scattering equivalent.  At first glance it looks like $V_a$ and $V_b$ 
are related by a simple variable change.  This is not the case because 
$V_a$ and $V_b$ commute with different kinematic subgroups.  This 
property cannot be changed by a change of variables. 

It is apparent from equations (\ref{b.5}-\ref{b.6}) that both 
representations have 
identical internal wave functions.  The relevant unitary transformation
is
\beq
\sumint{}_{\lambda j} A^{\lambda j}(\mathbf{v}_a;\mathbf{v}_b)
\vert \psi_{\lambda,j}\rangle \langle \psi_{\lambda,j} \vert  
\label{b.7}
\eeq
where this involves a sum over the eigenvalues $\lambda$ of the 
internal mass operator. 

To establish that both models give the 
same $S$ matrices we first observe that 
the structure of the interactions implies that the 
scattering matrices in both representations are 
\beq
\langle (m,j), \mathbf{v}_a, \mathbf{d} \vert S_a 
\vert  (m',j'), \mathbf{v}'_a, \mathbf{d}' \rangle =
\delta (\mathbf{v}_a,\mathbf{v}_a')\delta_{jj'} \delta (m,m') 
\langle \mathbf{d} \Vert S^{mj} \Vert \mathbf{d}' \rangle 
\label{b.8}
\eeq
and
\beq
\langle (m,j), \mathbf{v}_b, \mathbf{d} \vert S_b 
\vert (m',j'), \mathbf{v}'_b, \mathbf{d}' \rangle =
\delta (\mathbf{v}_b,\mathbf{v}_b')\delta_{jj'} \delta (m,m') 
\langle \mathbf{d} \Vert S^{mj} \Vert \mathbf{d}' \rangle .
\label{b.9}
\eeq
where the reduced $S$-matrices $\langle \mathbf{d} \Vert S^{mj} \Vert \mathbf{d}' \rangle$ are identical.

If we change variables in the first of these equations we get
\[
\langle (m,j), \mathbf{v}_b, \mathbf{d} \vert S_a 
\vert  (m',j'), \mathbf{v}'_b, \mathbf{d}' \rangle =
\]
\[
\sumint A^{m j}(\mathbf{v}_b;\mathbf{v}_a) d\mathbf{v}_a
\langle (m,j), \mathbf{v}_a, \mathbf{d} \vert S_a 
\vert \langle (m',j'), \mathbf{v}'_a, \mathbf{d}' \rangle
d\mathbf{v}_a'  
A^{m' j'}(\mathbf{v}_a';\mathbf{v}_b') =
\]
\[
\sumint A^{m j}(\mathbf{v}_b;\mathbf{v}_a) d\mathbf{v}_a
\delta (\mathbf{v}_a,\mathbf{v}_a')\delta_{jj'} \delta (m,m')
d\mathbf{v}_a'
A^{m' j'}(\mathbf{v}_a';\mathbf{v}_b') 
\langle \mathbf{d} \Vert S^{mj} \Vert \mathbf{d}' \rangle 
=
\]
\[
\delta (\mathbf{v}_b,\mathbf{v}_b')\delta_{jj'} \delta (m,m') 
\langle \mathbf{d} \Vert S^{mj} \Vert \mathbf{d}' \rangle 
=
\]
\beq
\langle (m,j), \mathbf{v}_b, \mathbf{d} \vert S_b 
\vert (m',j'), \mathbf{v}'_a, \mathbf{d}' \rangle 
\label{b.10}
\eeq
which proves the equivalence.

\section{Summary}

To summarize, Bakamjian-Thomas 
constructions of dynamical representations of the Poincar\'e group
have the general form
\beq
U_b(\Lambda ,a) \vert  (\lambda,j), \mathbf{v}_b  \rangle
= \sumint d\mathbf{v}' \vert  (\lambda,j), \mathbf{v}_b'  \rangle
 {\cal D}^{\lambda' ,j}_{\mathbf{v}_b',\mathbf{v}_b}[\Lambda ,a]
\label{c.11}
\eeq
where
\beq
\langle (m,j), \mathbf{v}_b \vert  (\lambda,j'), \mathbf{v}'_b  \rangle
=
\delta (\mathbf{v}_b,\mathbf{v}_b')\psi_{\lambda ,j} (m,\mathbf{d})
\label{c.12}
\eeq
and  $\psi_{\lambda ,j} (m,\mathbf{d})$ is the solution of
the mass eigenvalue equation: 
\beq 
(\lambda -m) \psi_{\lambda' ,j'} (m,\mathbf{d}) = \sumint{}'
\langle m, \mathbf{d} \Vert V^j \Vert m', \mathbf{d}'\rangle dm'
d\mathbf{d}' \psi_{\lambda' ,j'} (m',\mathbf{d}')  
\label{c.13}
\eeq
which is identical in all forms of dynamics.  Equivalent models with
different kinematic symmetries differ only in the choice of the
variables $\mathbf{v}_b$ in equations (\ref{c.11}-\ref{c.12}).  While
different choices of $\mathbf{v}_b$ lead to different interactions 
(\ref{a.2}) with different kinematic symmetries, the
resulting dynamical models are all equivalent.

Irreducible vectors in the different forms of dynamics are related by
\beq
\vert  (\lambda,j), \mathbf{v}_b  \rangle =
\sumint' 
\vert  (\lambda,j), \mathbf{v}'_c  \rangle d\mathbf{v}_c'
A^{\lambda j}(\mathbf{v}'_{c};\mathbf{v}'_{b})
\label{c.14}
\eeq
and the Wigner functions in different representations are related by 
\beq
{\cal D}^{\lambda ,j}_{\mathbf{v}_c',\mathbf{v}_c}[\Lambda ,a] =
\sumint d\mathbf{v}_{b} d\mathbf{v}'_{b}
A^{\lambda j}(\mathbf{v}'_{c};\mathbf{v}'_{b})
{\cal D}^{\lambda ,j}_{\mathbf{v}_b',\mathbf{v}_b}[\Lambda ,a]
A^{\lambda j}(\mathbf{v}_{b'};\mathbf{v}_{c}).
\label{c.15}
\eeq
The transformation relating the different kinematic subgroups are
dynamical because the mass eigenvalues $\lambda$ that appear in both
$A^{\lambda j}(\mathbf{v}_{b};\mathbf{v}_{c})$ and ${\cal D}^{\lambda
  ,j}_{\mathbf{v}_b',\mathbf{v}_b}[\Lambda ,a]$ are determined by
solving the dynamical equation.  The important observation is that the
physical observables (binding energies, S-matrix elements) are
obtained by solving (\ref{c.13}) which is independent of the choice of
kinematic subgroup.  

The conclusion of this work is that Poincar\'e invariant
quantum models should be considered as being defined without reference
to any specific kinematic subgroup, and any Poincar\'e invariant model
can be transformed to a representation that exhibits any kinematic
symmetry.  This conclusion is not limited to two-body models or models
that conserve particle number - nor is it limited to the maximal
kinematic subgroups discussed by Dirac.  The non-trivial 
dynamical equation that must be solved is the same in all cases.
The different choices of representation have no effect on bound state
or scattering observables.

The one class of applications where using different forms of dynamics
has dynamical consequences is when they are used in the one
photon-exchange approximation.  This is because the initial and final
hadronic states are in different frames, and have different invariant
masses.  The equivalence proof breaks down when $m\not= m'$.  While
the equivalence can be recovered by transforming the impulse current
in one representation to another representation, the transformed
current will generally have many-body contributions.

This work supported in part by the U.S. Department of Energy, under 
contract DE-FG02-86ER40286.
    
\bibliography{lf_ins}

\end{document}